\documentclass[a4paper,amsmath,amssymb,prl,twocolumn]{revtex4}

\usepackage{graphicx}
\usepackage{dcolumn}
\usepackage{bm}
\usepackage{epic,eepic}
\usepackage{graphicx}
\usepackage{subfigure}
\usepackage{psfrag}

\begin{document}
\author{Paolo {De Gregorio}, Aonghus Lawlor, Phil Bradley, Kenneth A. Dawson}

\address{
\mbox{Irish Centre for Colloid Science and Biomaterials,
Department of Chemistry,
University College Dublin,
Belfield,
Dublin 4,
Ireland}}

\title{Clarification of the Bootstrap Percolation Paradox}

\begin{abstract}
We study the onset of the bootstrap percolation transition as a model
of generalized dynamical arrest. We develop a new importance-sampling
procedure in simulation, based on rare events around ``holes'', that
enables us to access bootstrap lengths beyond those previously
studied. By framing a new theory in terms of paths or processes that
lead to emptying of the lattice we are able to develop systematic
corrections to the existing theory, and compare them to
simulations. Thereby, for the first time in the literature, it is
possible to obtain credible comparisons between theory and simulation
in the accessible density range.
\end{abstract}

\maketitle

The bootstrap percolation
\cite{adler1991,adler2003}
problem has attracted considerable interest from a variety of
scientific communities. In essence it (or its obvious variants) is a
method to analyze the dynamics of a system of highly coupled units,
each of which has a state that depends on those of its close
neighbors. Such units have been considered particles, processors, or
elements of a growing population. Units that become active in the
underlying dynamics are simply removed in the bootstrap. The ensemble
can undergo a transition to an ``arrested'' state in which all
dynamics is quenched, and this is (after bootstrap removal of units)
reflected in a transition from an empty to partially filled
lattice. This arrest is found to be driven by a long length scale. The
change of state of one unit becomes increasingly difficult near
arrest, requiring a long sequential string of favorable changes in its
surrounding units.  Thereby a long length (the size of the surrounding
region to be changed) is slaved to a long (relaxational) time scale.

The simplicity of bootstrap concepts means that bootstrap
percolation plays a canonical role in the conceptual framework of
the dynamical arrest transition, ``glassification''
\cite{jackle2002,sabhapandit2002,toninelli2003}
and in arenas as diverse as processor arrays
\cite{kirkpatrick2002} and crack propagation \cite{adler1988}
amongst others. 

We study two types of ``dynamics'', the well-known ``bootstrap model''
itself
\cite{chalupa1981,kogut1981,adler1991,manna1998}, and the modified bootstrap
\cite{adler1988,schonmann1992}. In the former, particles
are removed if they are surrounded by $c$ or less neighbors, and in
the latter they are removed if any two of its vacant nearest neighbors
are also second neighbors to each other. The (random bootstrap)
transition occurs as a function of $c$, system size $L$, and initial
particle density $\rho$ and occurs when half of the prepared initial
states are empty after removal of all movable particles.

In this paper we introduce both new simulation algorithms and
theoretical approaches that qualitatively change the regimes that
may be explored. The theory is now relevant to physically
accessible length-scales and, using only a personal computer, the
simulations can be extended beyond the largest scales currently
accessible in the most advanced simulations. A most interesting
outcome, perhaps of topical interest \cite{gray-review}, is that
we are able to elucidate the origin of disagreements between
simulation \cite{adler1988,kurtsiefer2003} and theory
\cite{aizenman1988,vanenter1988,holroyd2003} for the bootstrap models, and show
how the two can work together more closely in future
developments. Currently our calculations are detailed and specific to
these models, but we consider they contain the kernel of generality
required to signpost the path to future developments in the whole
arena of dynamical arrest.

The bootstrap-type problems mentioned above fall into two broad
``universality'' classes of arrest transition \cite{adler1991}. The
first type is a continuous or ``critical'' point transition in which
progressively more particles lose motion, leading to growing arrested
domains whose typical size diverges with a power law. The second type
of transition (of interest to us here), is more reminiscent of a
first-order transition. There, dynamically slowed domains grow near
arrest according to an essential singularity. Mobilization of these
domains is dependent on rare events (we call these ``connected holes''
\cite{lawlor2002prl}) involving specific units that can nucleate
motion on this large length. As will become clear, these nuclei become
highly dilute near the transition, the typical distance between them
being the diverging bootstrap length.

For such transitions the following conclusions have been drawn by the
community. For $c=d$ (the dimension) it is believed that the bootstrap
length $\xi$ diverges according to an essential singularity $\xi^{d} =
\exp ^{\circ(d)} (-A/(1-\rho))$ where $\exp^{\circ (d)}$ is the
exponential function iterated $d$ times \cite{adler1991,cerf1999}.
For the two dimensional square lattice $c=2$, theoretical calculations
\cite{holroyd2003} have resulted in an elegant outcome; essentially
what are believed to be exact results, $\lim_{ \stackrel{\rho \to
1}{\xi \to \infty} } 2 (1-\rho) \log
\xi=A$, where $A=\pi^{2}/9$ and $\pi^{2}/3$ for conventional and
modified bootstrap respectively.

On the other hand, all attempts to obtain this asymptotic result
by simulation have so far failed, including
extensive calculations
up to $L=128,000$, leading to speculation that it may be relevant to
particle densities very close to unity, and consequently system sizes
that are incredibly large (eg. $\xi \sim 10^{47}$ at $\rho=0.995$
\cite{prl_bootstrap_note2}). Such lengths would be far beyond what
will ever be possible, or indeed of interest, for Physics to explore.

Buried within this problem, however, a more troubling implication
emerges \cite{adler2003,holroyd2003,gray-review} that simulations
and theory seem relevant to such different scales that there can
be little useful dialogue between them. The point is also well
made that many of the interesting arenas of application involve
large, but finite number of units, and the present theory does not
seem helpful there \cite{adler2003}. We will show that this need
not be the case. When the problem is properly placed in context of
general knowledge in the Physics community, the previous theory is
useful, elegant and worth developing further, and the simulations
were correct, and worthwhile.

\begin{figure}[b]
\psfrag{rho}{{\Huge $\rho$}}
\psfrag{logp}{{\Huge $\log_{10}(\nu)$}}
\includegraphics[angle=-90,width=\columnwidth,height=!]{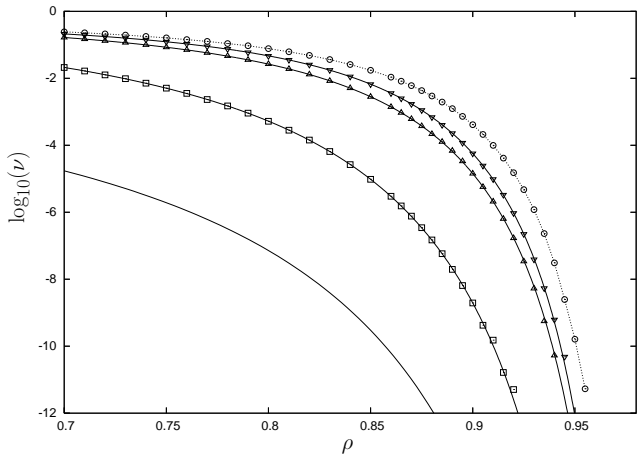}

\caption{Modified Bootstrap Model. The ($\circ$) points represent the
total hole density (the dotted line is a guide to the eye). The
symmetrically growing squares ($\square$) are compared with the
asymptotic result $\exp(-\pi^{2}/3(1-\rho))$ (lower solid line)
\cite{holroyd2003}. Also shown are results for diffusing squares
($\triangle$) and small-asymmetry rectangles ($\triangledown$)- in
these cases, the lines through the points represent our theoretical
results.}
\label{figure_modified}
\end{figure}
\begin{figure}[b]
\psfrag{rho}{{\Huge $\rho$}}
\psfrag{logp}{{\Huge $\log_{10}(\nu)$}}
\includegraphics[angle=-90,width=\columnwidth,height=!]{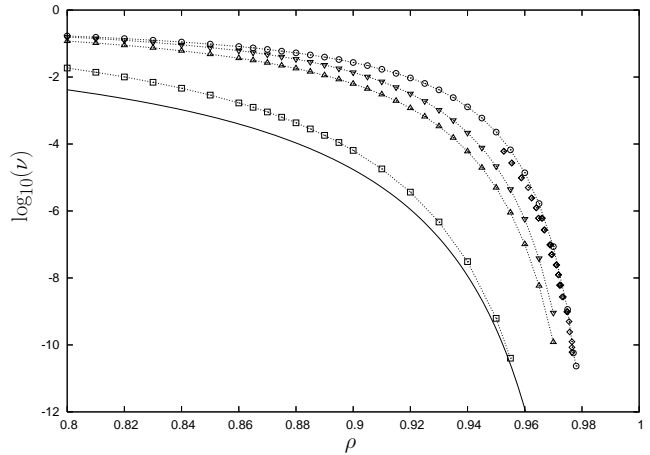}

\caption{Bootstrap Model. The ($\circ$) points represent the total
hole density (the dotted line is a guide to the eye) and can be
compared with the results for {$1/\xi^{2}$} ($\diamond$) for $\xi>100$
\cite{adler2003,stauffer_pers}. The symmetrically growing squares
($\square$) are compared with the asymptotic result $\nu =
\exp(-\pi^{2}/9(1-\rho))$ (lower solid line) \cite{holroyd2003}. Also
shown are results for diffusing squares ($\triangle$) and
small-asymmetry rectangles ($\triangledown$).}
\label{figure_bootstrap}
\end{figure}

We will show that the modified theory and simulation can be brought
into agreement over all reasonable length scales of interest to
Physics. Our results are exact also in the true asymptotic regime
\cite{holroyd2003} but simulations cannot reach there and Physics will
likely not be interested in the outcome.
We identify ``holes'' on the lattice as spaces (vacancies) into which
particles can move \cite{lawlor2002prl}. We then identify these holes
as either ``connected'' or caged (disconnected) according to whether
the lattice can (or cannot) be vacated by sequentially removing
particles beginning from that hole. The relationship to conventional
(random) bootstrap simulations (described above) is clear; a given
system size and density must contain at least one connected hole for
it to be vacated by random bootstrapping processes. Thus, the
bootstrap correlation length $\xi$ is related to the connected hole
density $\nu$ via $\nu=1/\xi^{2}$. The bootstrap length is therefore
the average distance between connected holes (these representing the
growth ``nuclei'' alluded to in our introductory remarks) that become
increasingly rare near arrest. The device of holes allows us to focus
on the key ``order'' parameter, rather than the very populous, but
irrelevant particles and vacancies \cite{lawlor2002prl}.

The details will be presented elsewhere, so here we present only
some results for the case $d=2$. In the simulations we begin by
creating a hole with the appropriate weight, and then populate
('grow') the configuration with particles and vacancies around
this site, checking at each stage to see if that hole is
connected, or trapped (ie a rattler \cite{lawlor2002prl}) by
identifying cages at that length. Since the typical cage size
grows much more slowly ($\log(1-\rho)/\log(\rho)$) than the
bootstrap length, we need check only relatively small distances
before the transition probability of bootstrap to the next largest
length approaches unity, and the hole is connected with certainty.
This approach has significant practical advantages since it
permits us to sample directly only the important rare events
(connected holes) rather than prepare and study very large
systems. Thus, the results produced here require only a few hours
of time on a personal computer. In Figure \ref{figure_modified}
and Figure \ref{figure_bootstrap} we show results for the total
connected hole density in the modified and conventional bootstrap
model, and these agree, where comparisons are available, with the
most extensive conventional simulations. For example, the
($\diamond$) points in the uppermost curve of Figure
\ref{figure_bootstrap}) represent the hole density implied by the
results in \cite{adler2003,stauffer_pers} (system size $L=128,000$),
while the ($\circ$) points on that same curve are from our
importance-sampling procedure discussed above. An additional advantage
in the simulation approach is that we can make direct contact with
theory.

Previous theoretical calculations
\cite{jackle2002,schonmann1992,holroyd2003} approximate the
process of simultaneous removal of particles on increasingly large
boundary contours until reaching one that is entirely occupied by
particles, and therefore immovable. The theory presented here is
developed by finding the most important 'paths' or sequences by
which particles are removed, counting the number of holes that
would be connected using such a set of paths. Addition of new path
types systematically improves the result, making comparison to
theory feasible for the first time.

Schematically, the probability of bootstrap $\nu = P_{\infty}$ is
represented $P_{\infty} = \Pi_{k=1}^{\infty} (1-\rho^{ak})^b$, where
$b$ represents the number of sides and $a$ the increment on the
lattice. In the high density limit one takes the natural logarithm of
the product, approximates the sum with an integral and makes the
substitution $y=\rho^k$. For $\rho \to 1$ this leads to
$-b/(a\ln\rho)\int_0^{1}(dy/y)\ln(1-y)\sim-b\pi^2/6a(1-\rho)$
\cite{prl_bootstrap_note2}. In the modified bootstrap model, $a=2$
and $b=4$, one obtains $-\pi^2/3(1-\rho)$, and it is this result that
has been proven to be asymptotically exact \cite{holroyd2003}. From
Figure \ref{figure_modified} it is clear (as has often been reported)
that there is no agreement between the simulated results for the total
hole density ($\circ$) and the asymptotic result, even for the highest
densities we can simulate. It is important to ask how much of this
deviation comes from the limited set of paths included in the theory
and how much from the fact that simulations can never reach the truly
asymptotic limit.
Thus, we calculate exactly the probability for removal of concentric
squares of particles, a result that includes all corner contributions,
and is valid for all densities. Then,
\begin{eqnarray}
&&P^{(cs)}_\infty=(1-\rho)\Pi_{k=1}^{\infty}c_k^{(cs)}(\rho) \label{1} \\
&&c_k^{(cs)}(\rho) = 1-4\rho^{2k+1}+2\rho^{4k+1}(2+\rho)-4\rho^{6k+1}+\rho^{8k} \nonumber
\label{eq_prob_squares}
\end{eqnarray}
Using $(1-\rho^{2k+1})^4$ as lower bounds for the coefficients
$c_k^{(cs)}(\rho)$ and we find a modified asymptotic result for the
hole density $(1-\rho)^{-5}\exp(-A/(1-\rho))$
\cite{prl_bootstrap_note2}. This modified asymptotic result is almost
equal to our numerical solution of Equation \ref{eq_prob_squares} and
simulation results for the symmetrically growing squares process in
the density range of interest. While these are in perfect agreement
with each other (see ($\square$) in Figure
\ref{figure_modified}), they are still many orders of magnitude
different from the full simulated connected hole density, although
there is clearly a considerable improvement over the purely asymptotic
result $\exp(-\pi^{2}/3(1-\rho))$.

Nevertheless, by permitting only such restricted paths we have
failed to correctly identify many holes as connected. Our aim now
in development of the theory is therefore to systematically
enlarge the possible paths in the calculation, until we approach
the full simulated result, at each stage validating the
theoretical calculation by simulating the same restricted set of
paths. By simulation it is easy to show that indeed paths
involving symmetric removal are the most probable, but the
constraint that all particles be removed from the boundary in a
single step results in loss of many holes. If instead the boundary
particles are simultaneously removed only on adjacent sides of the
square, then the emptying squares are also permitted to
``diffuse'', that is, change their center of gravity during the
process. This leads us to identify many new holes that would
otherwise be wrongly identified as disconnected. One can go one
stage further, and permit even asymmetric paths, eventually
exhausting all paths. However, by binning the paths adopted in the
simulation of the total connected hole density we find that
throughout the whole emptying process, beginning from a connected
hole, the ratio of the rectangular sides never exceeds $1.4$, by
far the greatest contribution coming from near-square process,
providing they are permitted to diffuse.
\begin{figure}[b]
\subfigure[$P^{[1]}$]{\includegraphics[width=0.19\columnwidth,height=!]{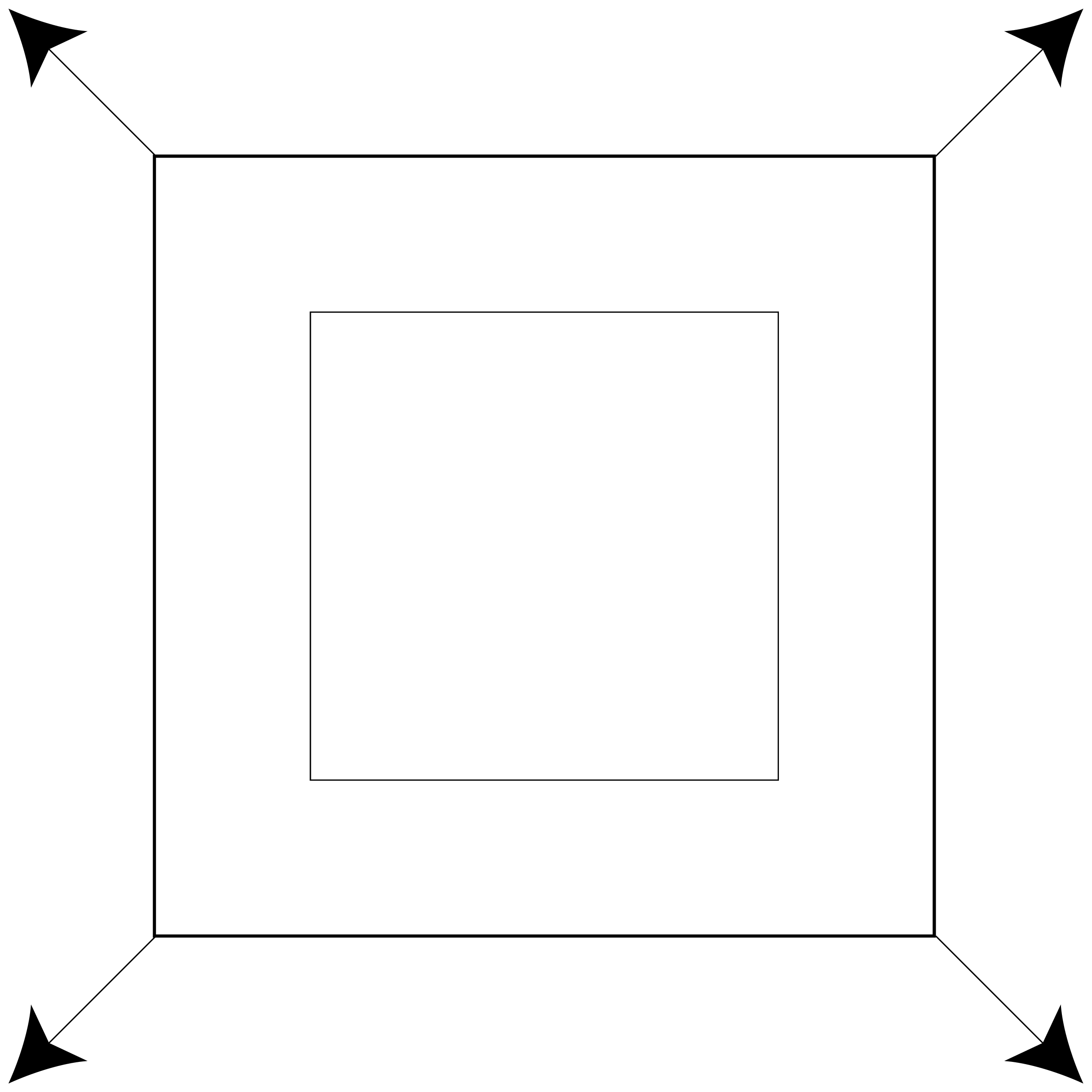}}\hfill
\subfigure[$P^{[2]}$]{\includegraphics[width=0.19\columnwidth,height=!]{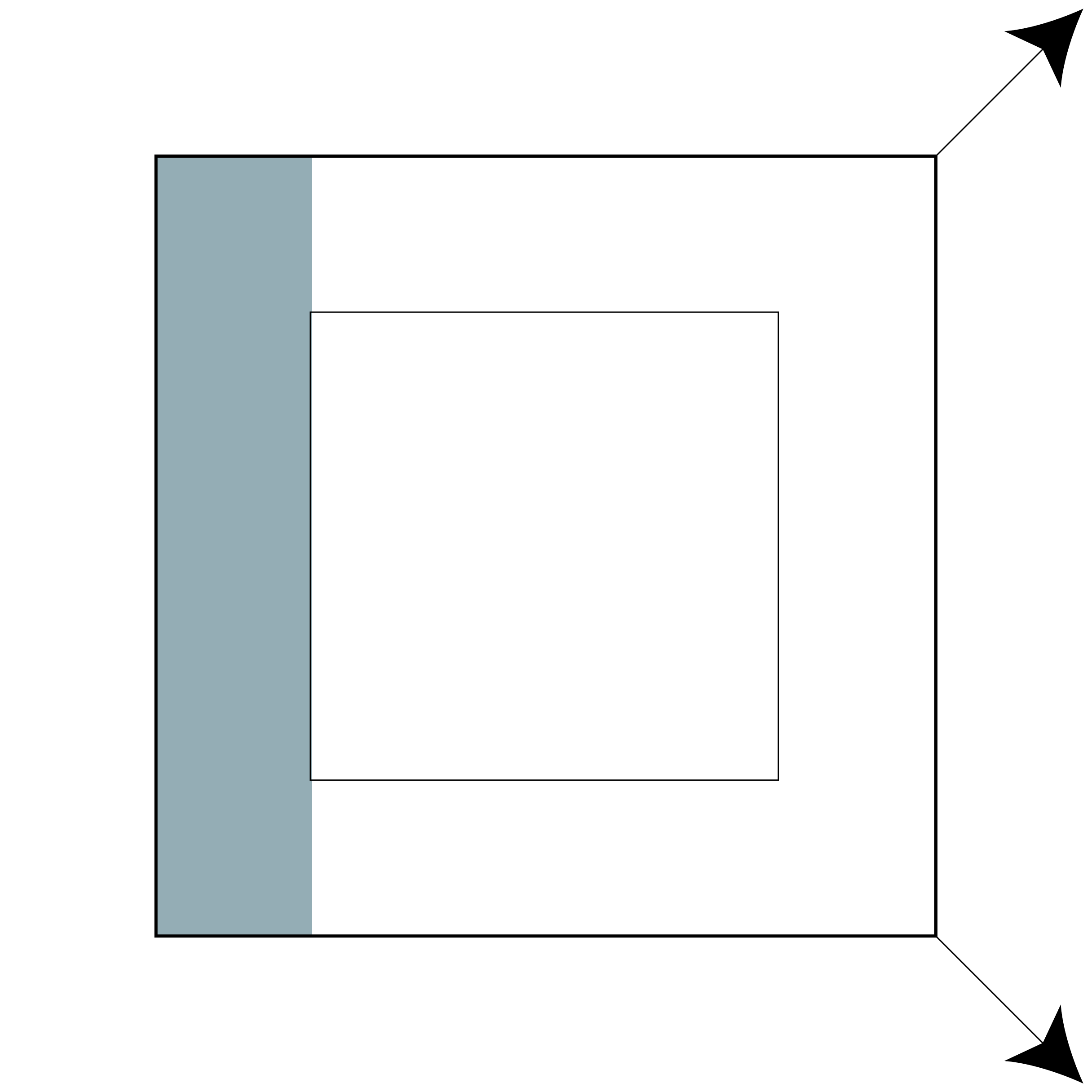}}\hfill
\subfigure[$P^{[3]}$]{\includegraphics[width=0.19\columnwidth,height=!]{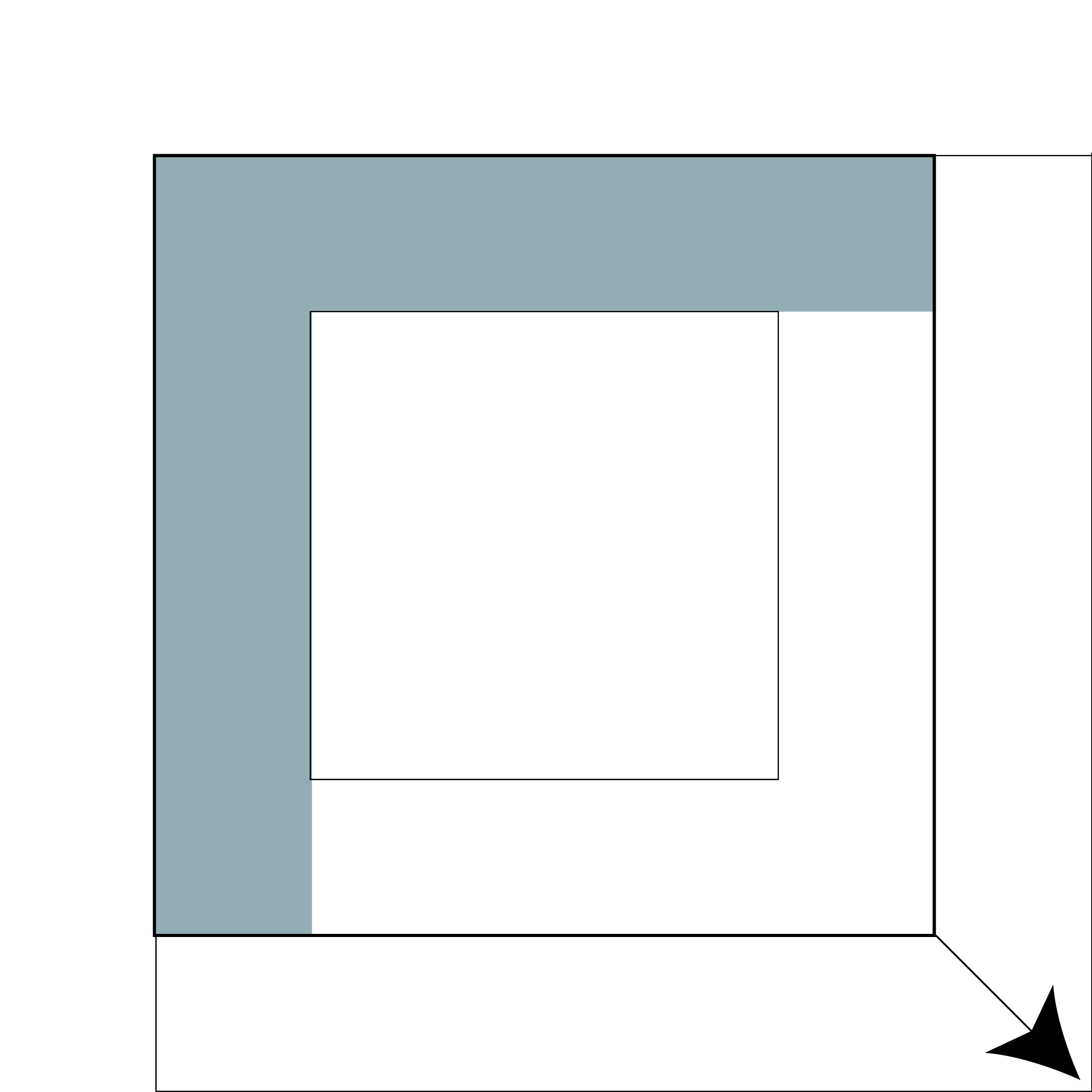}}\hfill
\subfigure[$P^{[4]}$]{\includegraphics[width=0.19\columnwidth,height=!]{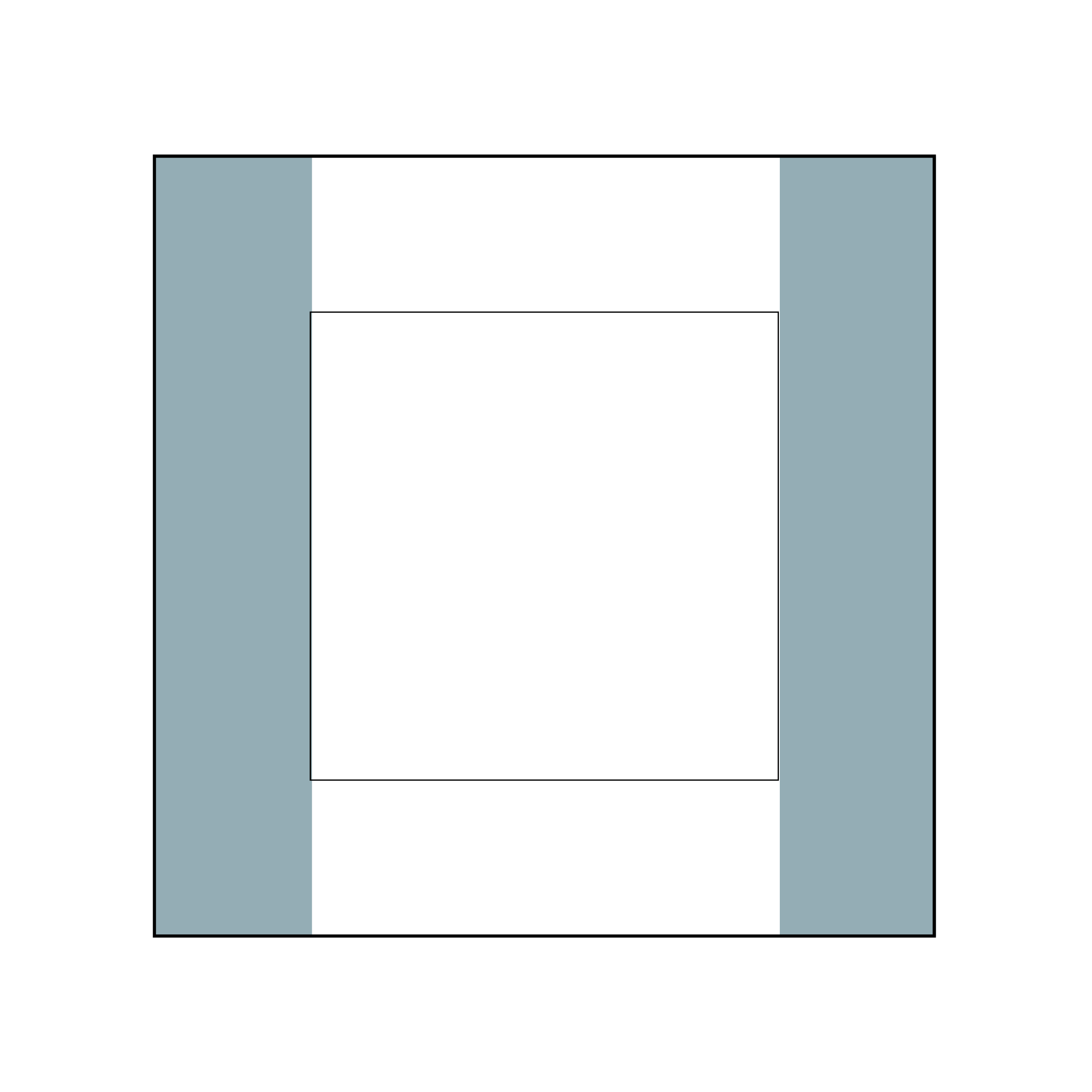}}\hfill
\subfigure[$P^{[5]}$]{\includegraphics[width=0.19\columnwidth,height=!]{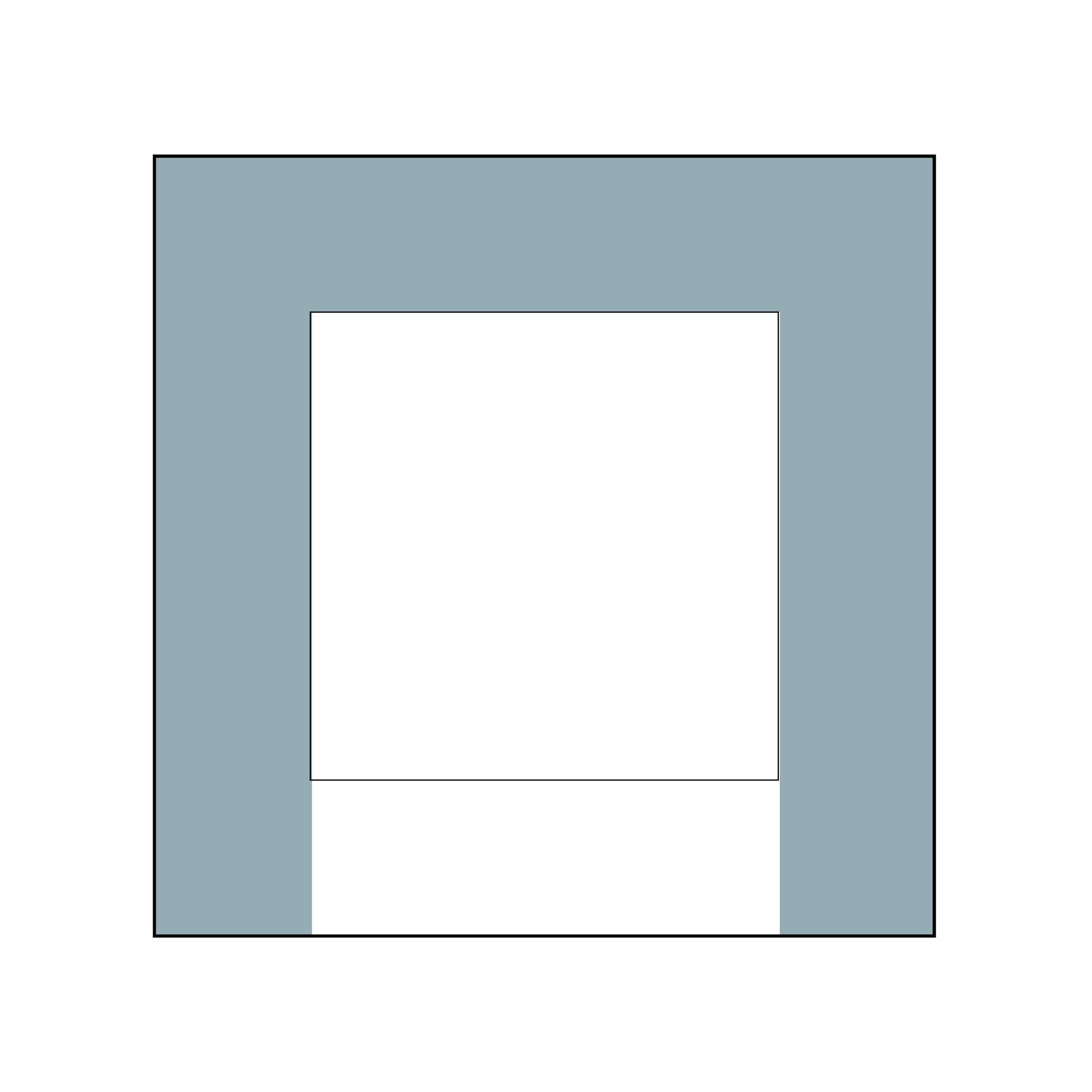}}\hfill
\caption{
Intermediate configurations arising in evaluation of the sum over
'paths' that may be used to empty the lattice. The inner square from
each of these figures is assumed to have been emptied of particles in
the previous step, and the next parts to be emptied lie between this
and the outer bounding square. Shaded outer regions imply a complete
line of particles, blocking further movement in that
direction. Unshaded outer regions have at least one vacancy in that
part of the perimeter. The arrows on the squares indicate possible
transitions involving the growth of two adjacent boundary lines by one
step. This process restores that local configuration to one of the
intermediate states again, with the internal empty square one step
larger in two directions. The process terminates when the local
configuration makes a transition outside of the class $P^{1}$ -
$P^{3}$ (diffusing squares) or $P^{1}$ - $P^{5}$ for small-asymmetry
rectangles. Note that in $P^{3}$ we have given an explicit example of
the extension of the boundary implied by the arrow.}

\label{figure_probs}
\end{figure}
We have been able to realize this approximation in theory also. We
define intermediate states of the system as the squares illustrated in
Figure \ref{figure_probs}, with weight $P_k^{(i)}$. Paths implied by
removal of particles map the growing vacant region between only these
states, larger by one step at each stage. If we consider only the
limited set $P^{1}$ - $P^{3}$, such processes correspond to growing
and diffusing squares. Inclusion of the states $P^{4}$ and $P^{5}$
permits in addition ``fluctuations'' of the square by one additional
layer. These local intermediate states are related by the coupled
equations,
\begin{eqnarray}
&&P_k^{(i)}(\rho)=\sum_jc^{(i,j)}_k(\rho)P_{k-1}^{(j)}(\rho)
\end{eqnarray}
where $i$, $j$'s range from $1 \to n$, with $n=3$ in the diffusing
squares process, and $n=5$ in its extended small-asymmetric
rectangular version. $c^{(i,j)}_k(\rho)$ defines the probability of
migration from class $j$ to $i$ at the $k$th step. These equations are
solved subject to the initial conditions,
$P_1^{(i)}=(1-\rho)\delta_{1i}$.

The choice of these states, and transitions between them, is far from
trivial since we must ensure an ``ordering'' of the removal process if
we wish to use a random measure for the particles in calculating the
coefficients $c^{(i,j)}_k(\rho)$. For the sake of simplicity, we here
present only the coefficients of the process involving diffusing
squares,
\begin{eqnarray}
&&c^{(1,1)}_k=1-2\rho^{2k}+\rho^{2k-1}, \hfill c^{(1,2)}_k=(1-\rho)(1-\rho^k) \nonumber \\
&&c^{(1,3)}_k=(1-\rho)^2, \qquad  c^{(2,1)}_k=2\rho^k(1-\rho^{k-1}) \nonumber \\
&&c^{(2,2)}_k=1-2\rho^k+\rho^{2k-2}, \hfill c^{(2,3)}_k=2\rho(1-\rho-\rho^{k-1}+\rho^k) \nonumber \\
&&c^{(3,1)}_k=\rho^{2k-1}, \qquad c^{(3,2)}_k=\rho^k(1-\rho^{k-1}) \nonumber \\
&&c^{(3,3)}_k=\rho^2(1+2\rho^{k-2}-4\rho^{k-1}+\rho^{2k-3}) \nonumber
\end{eqnarray}

The total bootstrap probability $P_{\infty}^{(1)}(\rho)$, may be
calculated numerically for any density, limited only by the
precision of the computer. The same processes may be simulated on
the computer and are in each case identical with the theory.
Results for all are given in Figure \ref{figure_modified} and
Figure \ref{figure_bootstrap} for modified and conventional
bootstrap.

The outcome is intriguing. Diffusing squares ($\triangle$),
improves the comparison between theory and full simulation, and
small-asymmetry rectangles ($\triangledown$) yields results that
may (for the first time) begin to be credibly compared to computer
simulation of the full simulated hole density, and implicitly the
bootstrap correlation length in the regime where simulations can
be carried out.

We may now summarize our results both in relation to the bootstrap
problem, and in a more broad context. Firstly, for the bootstrap
problem itself, new computational and theoretical approaches have
enabled simulations and theory to be brought into reasonable (indeed
arbitrarily good) agreement across a wide range of density. The fact
is that both the theory and simulation do not adopt the very simple
asymptotic form that has been quoted in the literature until one
reaches densities and length scales that are beyond the natural
interest of physics. There nevertheless remain many areas of physics
where extended regimes of dynamic slowing, and near-arrest are of
great importance, and these can be dealt with by the methods described
here. Secondly, by properly identifying the most probable paths
(incidently thereby respecting the symmetry of the problem), and
developing theory as a sum over only these paths, we obtain a very
useful approximation, and systematic corrections around it. In essence
this amounts to a sort of ``mean-field'' approximation in the
path-integral (sum), more familiar in field theory as the optimal
instanton trajectory \cite{polyakov1987}. Inclusion of small asymmetry
is equivalent to the 'shape' fluctuations included in next to leading
order in such calculations. This is more than an analogy; the
bootstrap process of this discontinuous transition produces an
essential singularity precisely because, underlying it, is the physics
of complex activated processes.

In this second point, but from a broader perspective, we have linked
the whole bootstrap endeavor (and crucially those areas of physics for
which it is considered relevant) to an arena of physics that is
already somewhat explored, and opened the pathway to numerous
developments in the theory of dynamical arrest, many of which will
immediately suggest themselves to the reader.

\acknowledgements{The authors have benefited from ideas of G.
Biroli. We acknowledge interactions at various stages with A. van
Enter, S. Franz, A. Holroyd, M. Mezard, A. Robledo, F. Sciortino,
M. Sellitto, D. Stauffer, P. Tartaglia. The work is supported by HCM
and DASM.}


\begin{thebibliography}{21}
\bibitem[{prl()}]{prl_bootstrap_note2}
\bibinfo{note}{The calculations for growing concentric squares reduce to
  estimation of the integral $\int_0^{\rho}(dy/y)\ln(1-y)$ in the limit
  $\rho\rightarrow1$. A numerical estimate of $-(1-\rho)\sum_1^\infty
  \ln(1-\rho^k)$ shows that at $\rho=0.995$ there is still a $1.5\%$ deviation
  from $-\pi^2/6$. Adding the first correction to the integral in $(1-\rho)$
  gives $ -\pi^2/6 + \ln\rho\ln(1-\rho)$, already a good estimate at $0.95$.}
\end{thebibliography}

\end{document}